\documentclass[aps,pra,10pt,twocolumn,superscriptaddress,tightenlines,nofootinbib,longbibliography]{revtex4-1}

\usepackage[usenames,dvipsnames]{color}
\usepackage{colortbl}
\usepackage[table]{xcolor}
\usepackage{enumitem}
\usepackage{dcolumn}
\usepackage{graphicx} 
\usepackage{epstopdf} 
\usepackage{natbib}
\usepackage[colorlinks=true,linkcolor=blue,citecolor=blue,urlcolor=blue]{hyperref}
\usepackage{hypcap}
\usepackage{verbatim, float}
\usepackage{psfrag}
\usepackage[normalem]{ulem}
\usepackage{physics}		
\usepackage{siunitx}
\usepackage{marvosym}
\usepackage{dsfont}
\graphicspath{{./Figures/}}
\usepackage[commandnameprefix=ifneeded]{changes}

\usepackage{xcolor}
\usepackage{tcolorbox}

\usepackage{amsmath}
\usepackage{cleveref}

\providecommand{\moy}[1]{\langle #1 \rangle}

\newcommand{\optproblem}[1]{\underset{#1}{\text{max}}}
\renewcommand{\eqref}[1]{Eq.~(\ref{#1})}

\begin{document}

\title{Realistic Bell tests with homodyne measurements}

\author{Enky Oudot}
\thanks{These two authors contributed equally}
\affiliation{ICFO-Institut de Ciencies Fotoniques, The Barcelona Institute of Science and Technology, 
Av. Carl Friedrich Gauss 3, 08860 Castelldefels (Barcelona), Spain}

\author{Gaël Massé}
\thanks{These two authors contributed equally}
\affiliation{LAAS-CNRS, Toulouse, France}

\author{Xavier Valcarce}
\affiliation{Universit\'e Paris-Saclay, CEA, CNRS, Institut de physique th\'eorique, 91191, Gif-sur-Yvette, France}

\author{Antonio Acín}
\affiliation{ICFO-Institut de Ciencies Fotoniques, The Barcelona Institute of Science and Technology,
Av. Carl Friedrich Gauss 3, 08860 Castelldefels (Barcelona), Spain}

\begin{abstract}
We analyze Bell inequalities violations in photonic experiments for which the measurement apparatuses are restricted to homodyne measurements. Through numerical optimization of the Clauser-Horne-Shimony-Holt inequality over homodyne measurements and binning choices, we demonstrate large violations for states with a bounded number of photons. When considering states defined within qubit local subspaces of two Fock states, such as NOON states, a violation is observed solely within the qubit Fock space spanned by zero and two photons. For more generic states, large violations are obtained. Significant violations are observed even for states containing three photons locally and under realistic values of noise and losses. We propose concrete implementations to achieve such violations, opening new avenues for Bell experiments with homodyne detectors.
\end{abstract}

\maketitle

\clearpage

\section{Introduction}
\label{sec:Intro}

Bell inequalities serve as a powerful tool to demonstrate the incompatibility of quantum predictions with classical models in which measurements are pre-determined~\cite{bell1964,Brunner14}. 
Initially explored for its foundational implications, Bell nonlocality has now emerged as the core resource for quantum information  processing in a \textit{device-independent} (DI) manner, \textit{i.e.} without making assumptions about the underlying quantum model.
Notably, nonlocality has been leveraged to certify quantum resources, thanks to self-testing~\cite{Mayers2004,vsupic2020self}, and to provide device-independent security proofs for quantum key distribution~\cite{Acin2007, Pironio_2009, Vazirani2014, Sekatski2021, zapatero2019long, Schwonnek_2021, zapatero2023advances}. 

DI protocols require a  violation of a Bell inequality, with the Clauser-Horne-Shimony-Holt (CHSH) inequality being the most iconic~\cite{CHSH1969}.
 Bell tests have been successfully conducted~\cite{giustina2015significant,Hensen2015,shalm2015strong, Rosenfeld2017,Li2018, Storz2023}, and some DI protocols have seen their first realizations~\cite{Liu2018,Liu2022,Zhang2022,Nadlinger2022}.
However, so far these tests  involve measurements devices relying on superconducting nanowire single-photon detectors, which require cryogenic temperatures to operate efficiently.
This stands in contrast with the practical and commercial expectations for the implementation of DI protocols, where on-chip integration is highly desirable.

 We here seek to find set-ups that can achieves large CHSH violations with standard photonic devices. Thanks to their high efficiency, homodyne detectors are considered promising to close the detection loophole. Moreover, they fulfill two important criteria for practical Bell tests: being adapted to telecom wavelength and being able to include integrable detectors. Wenger \textit{et.al.} showed that one can find arbitrarily high violations of CHSH, based on continuous variables~\cite{Wenger2003}. This result was also generalised to the multipartite case in~\cite{Ferraro09}. Unfortunately, the states proposed to attain the maximal CHSH violation are not fitted for experimental implementations. On the other hand, the authors of \cite{GarciaPatron2004,GarciaPatron2005} proposed a setup, using photon subtraction on a two-mode squeezed state, that leads to a CHSH value of $S\approx 2.048$. While the required state is feasible in present experiments, the reported violation is fairly small and thus of limited use for DI applications. The situation is therefore as follows: too small violations have been obtained for realistic setups, while large, and even optimal violations are possible for states completely out of reach. This work reduces the gap between experimental feasibility and high violation of Bell inequalities.

In this manuscript, we derive Bell violations for the CHSH inequality using homodyne measurements. Notably, we found violations that are significantly larger than those found in previous works for realistic and robust experiments. After a brief reminder on Bell scenarios and on homodyne apparatuses in Sec.~\ref{sec:OptimHomo}, we explore violations of the CHSH inequality in local qubit Fock spaces in Sec.~\ref{sec:witness}. We then consider local qudit Fock spaces of growing dimensions. For each case, we give the CHSH score optimised over measurement parameters in Sec.~\ref{sec:growingdimensions}, and an analysis of the robustness considering a realistic noise model in Sec.~\ref{sec:noise}. Finally, in Sec.~\ref{sec:setups} we propose experimental set-ups that implement some of the violations derived in the precedent sections.

\section{Preliminaries}
\label{sec:OptimHomo}

\subsection{Bell scenarios}
\label{subsec:bellineq}

We consider a scenario in which two distant users, Alice and Bob receive correlated particles from a source. Each one can select locally a measurement indexed by $x,y$ over a choice of $m$ different ones. Each measurement can yield $\Delta$ possible outcomes~\cite{Brunner14} noted $a$ for Alice and $b$ for Bob. By repeating this protocol and sharing the results, it is possible to compute the conditional probabilities of obtaining outcomes knowing the local choices of measurements, written $P_{AB}(a,b|x,y)$. Probabilities compatible with a local hidden variable (LHV) model can be expressed as
\begin{equation}
\label{local}
P_{AB}^\textup{loc}(a,b|x,y) = \int_{\lambda} P(\lambda)P_A(a|\lambda,x) P_B(b|\lambda,y) \dd \lambda.
\end{equation}
Such distributions are called \emph{local}. A Bell inequality is an upper-bound on a linear combination of the probabilities $P_{AB}(a,b|x,y)$, given by the maximum value achievable with probability distributions of the form \eqref{local}.

\subsection{The CHSH inequality}
\label{subsec:chsh}
\begin{figure}
    \centering
    \includegraphics[width=1.\linewidth]{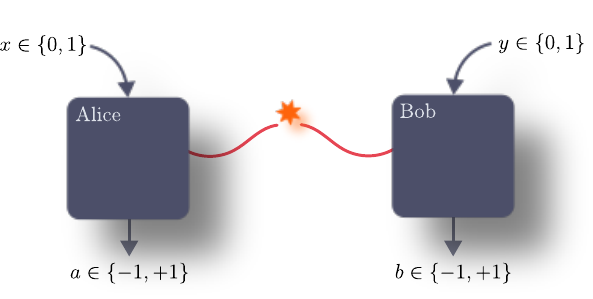}
    \caption{Schematic representation of a bipartite Bell test. In each round, Alice and Bob independently
chose a setting $x$ and $y$ and perform the corresponding measurement on the shared state. Then, they record their respective outcome $a$ and $b$. After many rounds, they compute the statistics $P_{AB}(ab|xy)$ from their choice of settings and outcomes.}
    \label{fig:biparitecorrexperiment}
\end{figure}

In this article, we consider bipartite Bell inequalities, with a focus on the simplest non trivial scenario. Each party can chose between two measurements, while the outcomes they obtain can take two values, either $-1$ or $1$, as depicted in Fig.~\ref{fig:biparitecorrexperiment}. In this case, there exists only one Bell inequality, the so-called Clauser-Horne-Shimony-Holt (CHSH) inequality \cite{CHSH1969} which reads
\begin{equation}
\label{CHSH}
     \moy{\mathcal{B}_\text{CHSH}} = \moy{A_0 B_0} + \moy{A_0 B_1} + \moy{A_1 B_0} - \moy{A_1 B1}\leq2.
\end{equation}
with $\moy{A_x B_y} = \sum_{a,b} p(a=b|x,y) - p( a \neq b |x,y)$, and where $A_i (B_j)$ stands for measurements by Alice (respectively Bob). The local bound of the CHSH inequality is $2$ whereas the maximum value predicted by quantum theory reaches $2 \sqrt{2}$~\cite{Tsirelson1987}. Note that in section~\ref{sec:growingdimensions} and~\ref{sec:noise}, we explore violations of other Bell inequalities, when considering that Alice and Bob have access to more than two measurement settings. 

\subsection{Bell test with homodyne measurement}
An homodyne measurement quantifies quadratures of the electromagnetic field. It corresponds to the measure of the operator 
\begin{equation}
    \hat{X}_{\theta}=\frac{\hat{a} e^{-i\theta}+\hat{a}^{\dagger}e^{i \theta}}{2},
\end{equation}
where $\hat{a}$ and $\hat{a}^\dagger$ are the ladder operators, and with $\theta\in[0,2\pi]$.
Note that the operator $\hat{X}_{\theta}$ has a spectrum in $\mathds{R}$. In order to step in the framework of a Bell test, a transformation from $\mathds{R}$ into $\{-1,1\}$ is necessary. This is called \emph{binning}. The positive operator-valued measure (POVM) elements corresponding to binned measurements read
\begin{equation}
    \begin{split}
        \{\prod\nolimits_{+1}^{\theta,I},&\prod\nolimits_{-1}^{\theta,I}\} \\
        &= \{\int_{I} \text{d}X_{\theta}\ket{X_{\theta}}\bra{X_{\theta}},\int_{\overline{I}} \text{d}X_{\theta}\ket{X_{\theta}}\bra{X_{\theta}}\}
    \end{split}
\label{eq:meas}
\end{equation}
where $\ket{X_{\theta}}$ is an eigenstate of the generalized quadrature $\hat{X}_{\theta}$, $I$ is a set of $\mathds{R}$ and $\overline{I}$ is its complementary. The observable corresponding to such POVM elements is
\begin{equation}
\label{localobs}
    \sigma^{\theta,I}=\prod\nolimits_{+1}^{\theta,I}-\prod\nolimits_{-1}^{\theta,I}.
\end{equation}
One can thus write the CHSH operator as
\begin{eqnarray}
    \label{eq:Belloperator}
    \nonumber
    \mathcal{B}_{\text{CHSH}}= \sigma_{A1}^{\theta_{A1},I_{A1}} \sigma_{B1}^{\theta_{B1},I_{B1}}+ \sigma_{A1}^{\theta_{A1},I_{A1}} \sigma_{B2}^{\theta_{B2},I_{B2}}\\
    + \sigma_{A2}^{\theta_{A2},I_{A2}} \sigma_{B1}^{\theta_{B1},I_{B1}}- \sigma_{A2}^{\theta_{A2},I_{A2}} \sigma_{B2}^{\theta_{B2},I_{B2}}
\end{eqnarray} 
where $\vec{\theta} = \{\theta_{A0},\theta_{A1},\theta_{B0},\theta_{B1}\}$ is the vector of the phase space direction of each measurement. 

\section{Violation of the CHSH inequality with homodyne measurement in local qubit Fock spaces}
\label{sec:witness}

In this section, we focus on the Hilbert qubit space $\mathcal{H}^{2}$ spanned by the Fock states $\{ \ket{l}, \ket{m} \}$. Theses basis represents states containing exactly $l$ and $m$ photons respectively. We start by extracting analytical lower bounds on the CHSH violation from the expression of the POVM elements. We then implement a numerical optimization to systematically explore higher possible CHSH violation.

\subsection{CHSH Score from the POVM Elements}

 Any 2-outcome qubit POVM can be written
\begin{equation}
\begin{split}
E_1 &= \mu \ket{\vec{n}}\bra{\vec{n}} + (1-\mu) r_1 \mathds{1}, \\
E_2 &= \mu \ket{\vec{-n}}\bra{\vec{-n}} + (1-\mu) r_2 \mathds{1},
\end{split}
\label{povm}
\end{equation} 
where $\mu \in \{0,1\}$ and $r_1=1-r_2 \in \{0,1\}$ quantifies the projective and the random part of the measurement respectively, and $\ket{\pm \vec{n}}$ are two orthogonal states in $\mathcal{H}^{2}$ (see Appendix~\ref{app:sec2outcome}). $\mu$ relates to the ability of the POVM to distinguish between two orthogonal states, which is a necessary condition to violate the CHSH inequality. Therefore, maximizing $\mu$ in POVM's expressions might lead to higher CHSH score. 

The projective part $\mu$ in direction $\ket{\pm \vec{n}}$ can be expressed following
\begin{equation}
\mu = \frac{1}{2} \left( \bra{+\vec{n}}E_1\ket{+\vec{n}} + \bra{-\vec{n}}E_2\ket{-\vec{n}} - 1 \right).
\end{equation}
Using the POVM expression for binned homodyne measurement defined in \eqref{eq:meas}, we have 
\begin{equation}
\begin{split}
\mu_h = \frac{1}{2} \bigg( &\int_{I} \text{d}X_{\theta}p(\theta,\ket{+\vec{n}}) \\ 
&+ \int_{\overline{I}} p(\theta,\ket{-\vec{n}}) \text{d}X_{\theta} - 1 \bigg)
\end{split}
\end{equation}
where $p(\theta,\ket{\pm\vec{n}}) = |\langle X_{\theta}|\pm \vec{n}\rangle|^2$. 
From this expression, we deduce that the maximum value of $\mu_h$ is reached when the set $I$ is composed of the intervals of $\mathds{R}$ for which the inequality $p(\theta,\ket{+\vec{n}}) > p(\theta,\ket{-\vec{n}})$ is satisfied. For this binning choice, $\mu_h$ can be computed from the trace distance between the two probability distributions $p(\theta,\ket{\pm\vec{n}})$ and $p(\theta,\ket{\pm\vec{n}})$. The maximum value of $\mu_h$ over the phase space direction $\theta$ and states $\ket{\vec{n}}$ is therefore given by
\begin{equation}
\max_{\theta,\ket{\vec{n}}} \frac{1}{2} \int \left\| p(\theta,\ket{+\vec{n}}) - p(\theta,\ket{-\vec{n}}) \right\| \text{d}X_\theta.
\label{eq:mumax}
\end{equation}
The maximum value of \eqref{eq:mumax} is achieved for $\theta=0$ and real orthogonal states $\ket{\pm n}$
\begin{eqnarray}
     \ket{+\vec{n}} &=& \cos(a)\ket{l} + \sin(a)\ket{m} \\
     \ket{-\vec{n}} &=& \cos(a)\ket{l} - \sin(a)\ket{m}
\end{eqnarray}
where $a \in [0,2\pi]$ (see Appendix \ref{app:sec2outcome}).

Moreover, we show in Appendix~\ref{app:sec2outcome} that in the case where Alice and Bob use the same binning, for all $\mu_h$ such that $|\cos(2a)^2| \leq \frac{1}{2}$, a CHSH score of 
\begin{equation}
    S = \mu_{h}^2 2\sqrt{2}
\end{equation}
can always be achieved.
Consequently, the task of optimizing the trace distance between two phase space probability distributions of two orthogonal states boils down to optimizing a function of only one real parameter $a$. We optimize the value of this CHSH $\text{score}$ for qubit Fock states $\{\ket{l}, \ket{m}\}$ going from $\ket{0}$ to $\ket{7}$. Surprisingly, the only space where we found a violation is the space spanned by $\{ \ket{0}, \ket{2} \}$. In this case, the maximization of the CHSH score can be performed analytically and leads to $S \approx 2.1477$, obtained for the state
\begin{equation}
    \ket{\psi} = \alpha \ket{00} + \beta \ket{02} + \beta \ket{20} + \alpha \ket{22},
    \label{eq:stateviolation}
\end{equation}
with $\alpha \approx -0.6504 - 0.0466 i$, $\beta \approx 0.0124 - 0.2514i$, and for a binning defined by  $I=[-0.8886,0.8886]$.

\subsection{Numerical optimization}
\label{subsec:numqubit}

We consider the eigenstate  $\ket{\psi}$ and  the eigenvalue $\lambda$  of the CHSH operator 
\begin{equation}
   \mathcal{B}_{\text{CHSH}}\ket{\psi} = \lambda \ket{\psi}.
\end{equation}
Since all quantum states can be decomposed in a basis of the eigenvectors of $\mathcal{B}_{\text{CHSH}}$, optimizing the CHSH score over all quantum states can be written
\begin{equation}
\max_{\psi} \bra{\psi}\mathcal{B}_{\text{CHSH}}\ket{\psi} = \lambda_{\max},    
\end{equation}
where $\lambda_{\max}$ is the highest eigenvalue of $\mathcal{B}_{\text{CHSH}}$. Using~\eqref{eq:Belloperator}, we deduce that $\lambda_{\max}$ depends on the choices of the binning $I$ and the phase space directions of measurements $\theta$ of Alice and Bob.  
For measurements $\{A_x,B_x\}$, we define the binning strategy for a set $I_x$ reading
\begin{equation}
    I_x = [a^x_0,a^x_1] \cup [a^x_2,a^x_3] \cup ... \cup [a^x_{q-1},a^x_q]
\end{equation}
with $\{a^x_0, a^x_1,...,a^x_q\}~=\vec{a}^{x}_q \in \mathds{R}^{q+1}$ .  We aim to find 
\begin{equation}
\max_{\vec{\theta},\vec{a}^{0}_q,\vec{a}^{1}_q} \text{Eig}(\mathcal{B}_{\text{CHSH}}).
    \label{eq:GenOptimbinning}
\end{equation}

 As the choice of $q$ directly impacts the optimization result, we performed optimizations over a growing number of binning elements until a similar score $S$ is reached for $q$ and $q+1$.

We run optimization \eqref{eq:GenOptimbinning} on qubit spaces spanned by Fock states $\{ \ket{l}, \ket{m} \}$ up to $l,m=20$.  In all cases except $\{ \ket{0}, \ket{2} \}$, the maximum score we obtain saturates the local bound. 
It is because a binning strategy such that the probability of getting the outcome $+1$ is $1$ leads to a CHSH score of $2$ in all cases.
This is consistent with the results derived in the previous section.
In the $\{ \ket{0}, \ket{2} \}$ case, compared to the analytical method, we slightly improved the score to $S \approx 2.1493$.
This improvement originates from a more refined binning choice, as different binning are allowed for measurements $A_0,B_0$ and $A_1,B_1$. These binning are given by $I_0=[-0.8886,0.8854]$ and $I_1=[-0.8689,0.8679]$ respectively.
\smallbreak
These two approaches strongly indicate that for qubit Fock spaces with up to $20$ photons locally, no CHSH violation using homodyne measurement can be found except for the qubit Fock space $\{\ket{0},\ket{2}\}.$

\section{CHSH score from homodyne measurements in qudit Fock space}
\label{sec:growingdimensions}

In this section, we extend the numerical optimization defined in Sec.~\ref{subsec:numqubit} to qudit Fock spaces.
We first compute optimized CHSH scores for states of local dimension up to 10 in Sec.~\ref{subsec:growingdimension} before restricting to energy conserving states in Sec.~\ref{subsec:energyconserving}.

\subsection{CHSH scores with homodyne measurements in local dimensions $3$ to $10$}
\label{subsec:growingdimension}

 For a local dimension $d$, we consider the Fock space spanned by the basis $\{\ket{0},....,\ket{d-1}\}$. The state shared between Alice and Bob is, therefore, a two-qudit state that lives in $\mathcal{H}^{d\times d}$. We run optimizations up to a maximum dimension of $d = 9$ as it corresponds to an upper bound of what can be possibly crafted with today's state of the art squeezing devices~\cite{Vahlbruch2016}.
 
\begin{figure}
    \centering
    \includegraphics[width= 0.95\columnwidth]{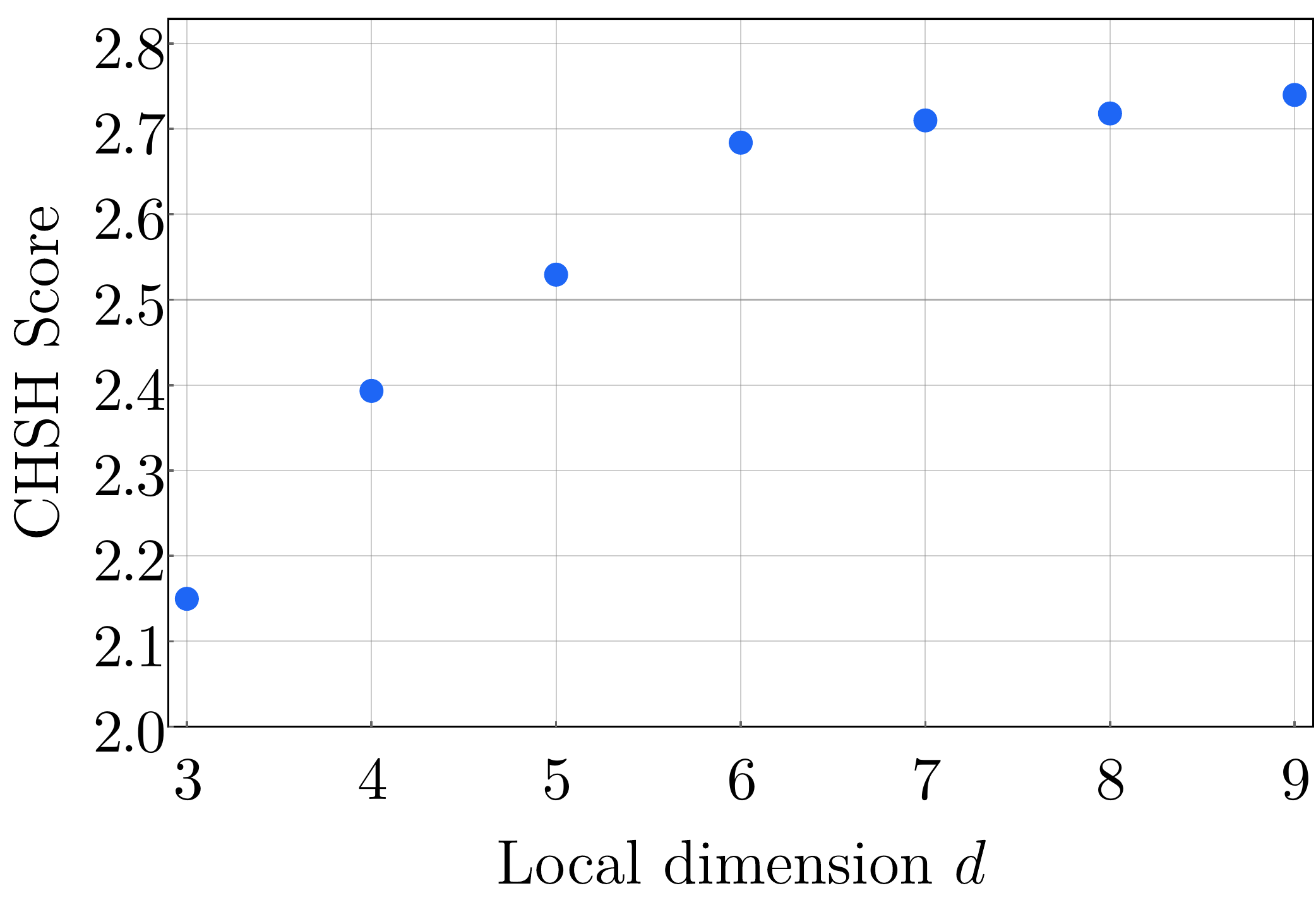}
      \caption{Maximum possible quantum values for the CHSH inequality as a function of the local dimension of the observables. These results are certified with a numerical observation. Note that in dimension $3$ we retrieve the score $\approx 2.14$ derived in Sec.~\ref{sec:witness}.}
    \label{fig:CHSHoptimbinningideal}
\end{figure}

The maximum CHSH scores we obtained for various local dimensions are given in Fig~\ref{fig:CHSHoptimbinningideal}. We observe that the larger the space dimension we consider, the larger the score. In local dimension $d=3$, we retrieve the result of the $\{\ket{0}, \ket{2} \}$ space, leading to a score of $S\approx 2.14$. In dimension $d=9$, we found a state yielding a score of $S\approx 2.7397$, close to the quantum bound.

\subsection{Quantum bounds for energy-conserving states}
\label{subsec:energyconserving}

We consider the specific case of optimizing the CHSH score in the case of energy-conserving states, due to their relevance for experimental implementation. In the case of a maximum number of $2$ photons, it means that the observables are $3 \times 3$ matrices in the basis $\{ \ket{02}, \ket{11}, \ket{20} \}$ ; with a maximum number of $3$ photons, the observables are $4 \times 4$ matrices in the basis $\{ \ket{03}, \ket{12}, \ket{21}, \ket{30} \}$, and so on and so forth. We verified that none of these constructions can yield a violation up to dimension $5$.

\section{Robustness with respect to noise}
\label{sec:noise}

Up until now we have considered ideal situations. However, in real-life, several factors generate so-called noise in experiments. For photonic experiment the leading source of noise is amplitude-damping.  In this section, we study the robustness of the CHSH test with respect to losses. 

We model photon losses by entangling an ideal incoming state with an ancillary fluctuating quantum field that we set to the vaccum state $\ket{0}$. After recombination on a beam-splitter, two outputs are produced corresponding to the transmitted part of the beam-splitter and to the reflected one. In order to obtain the noisy operator, we trace-out the part corresponding to the reflection. Hence, stemming from the ideal observable~\eqref{localobs}, we get the noisy observable
\begin{equation}
  \sigma_{A,\eta}^{\theta,I}  =\bra{0} \hat{U} \ \sigma_A^{\theta,I} \ \hat{U}^{\dagger} \ket{0}
    \label{eq:NoisyOp}
\end{equation}
where $\hat{U}$ is the beam-splitter observable defined by $ \hat{U} = e^{i \gamma (\hat{a}^\dagger \hat{b} - \hat{a} \hat{b}^\dagger)} $, $\hat{a}, \hat{b}$ are the annihilation operators respectively on the first and second mode, and the reflectivity of the beam splitter is $\eta = \cos(\gamma)^2$ (see Appendix~\ref{App:NoiseModel} for the derivation). In this way, we construct a new Bell operator $\mathcal{B}_{\text{CHSH}}^{\eta}$ by replacing $\sigma_{i}^{\theta,I}$ with $\sigma_{i,\eta}^{\theta,I}$ in \eqref{eq:Belloperator}.

We thus perform the optimization 
\begin{equation}
\max_{\vec{\theta},\vec{a}^{n}_i} \text{Eig}(B_{\text{CHSH}}^{\eta}).
    \label{eq:GenOptimNoise}
\end{equation}

\begin{figure}
    \centering
    \includegraphics[width=1.\linewidth]{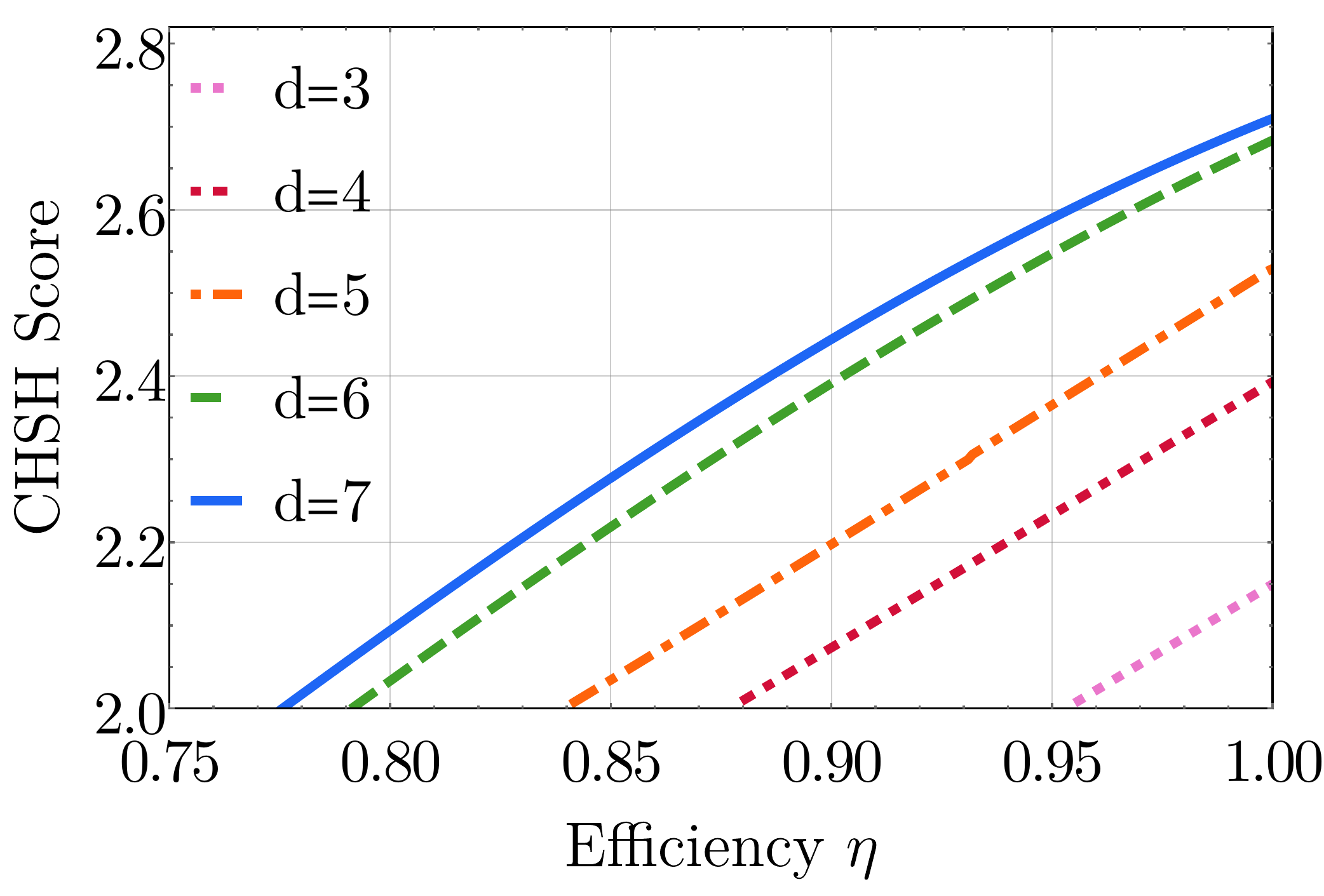}
    \caption{Result of the optimization~\eqref{eq:GenOptimNoise} yielding a CHSH score with respect to the overall efficiency of the protocol $\eta$. We plot the curves of the different ordered local Fock spaces $\{\ket{0},...,\ket{d}\}.$}
    \label{fig:noise}
\end{figure}

This allow us to obtain a threshold efficiency $\eta_c$ that depends on the incremental dimension of the space tested. We display the result of this optimization in Fig~\ref{fig:noise}. We see that by going to higher dimension one can increase the losses robustness of the CHSH violation. In the space spanned by $\{\ket{0},...,.\ket{6}\}$ we obtain a robustness of $\eta_c \approx 0.77$ for the losses.

Additionally, we explore the robustness to losses for scenarios where Alice and Bob have up to 4 local measurement settings.
All the 175 Bell inequalities of these scenarios can be found in \cite{Oudot_2019}.
We compute the optimisation~\eqref{eq:GenOptimNoise} for all of them.
Interestingly, for all tested local dimensions, no inequalities yield a better threshold efficiency $\eta_c$ than the CHSH inequality.

\section{Realistic implementation }
\label{sec:setups}

\begin{figure}[t]
	\begin{center}
		\includegraphics[width=0.45\textwidth]{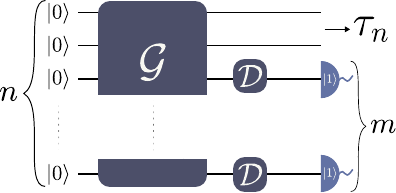}
	\end{center}
	\caption{$n$-mode bosonic parameterized circuits. The modes are initialized to the vacuum $\ket{0}$. A Gaussian process $\mathcal{G}$ is applied on all modes. $m=n-2$ displacement operations $\mathcal{D}$ are performed on the last modes, before heralding operations on single photon count. The state $\tau_n$ is then send to Alice and Bob.}
 \label{fig:circuit}
\end{figure}

To emphasize our results, we study the experimental implementation of quantum states $\rho_d$ yielding the highest CHSH score, for some of the maximum local dimensions constraints $d$.
We consider implementations based on Gaussian processes and heralding operations using photon-number resolving (PNR) detectors.
More specifically, we focus on $n$ bosonic modes circuits in which a $n$-mode Gaussian unitary transformation with zero-displacement, $\mathcal{G}$, is applied, followed by real displacements operations on every but the first two modes, see Appendix \ref{app:Circuits}.
The $m=2-n$ last modes are then heralded on single photon count, while the first two modes output a final state $\tau_n$ send to Alice and Bob. 
Fig.~\ref{fig:circuit} represents the photonic circuits under consideration.

Note that the Gaussian process $\mathcal{G}$ can be realisticaly implemented using an array of single mode squeezers followed by passive non linear optics, combining beam-splitters and phase shifters~\cite{Bloch1962,Braunstein2005,Gianfranco2016}. 
Moreover, recent results display successful implementations of heralding systems using PNR detectors~\cite{Davis2022,Stasi2023}.

\smallbreak

We optimize the fidelity $F(\rho_d, \tau_n)$ over the Gaussian process and the displacement operations.
This is achieved using the Riemannian optimization on the symplectic group described in \cite{Yao2022} and implemented in the \textsc{MrMustard} library~\cite{Yao2023Github}. 
Due to convergence instability with a higher number of modes $n$, we perform this optimization multiple times to avoid local minimum. 

We focus on target states $\rho_d$ with maximum local dimensions $d=\{3,4\}$. 
For each target state, we optimize circuits composed of $3$ to $7$ modes. 
In Appendix \ref{app:Circuits}, we provide the optimized fidelities and the corresponding squeezing and displacement parameters.
Importantly, these parameters are within the realm of experimental feasibility.

For $d=3$, the target state $\rho_3$ which achieves a CHSH score of $S\approx 2.14$ can be exactly prepared using a circuit with $n\geq 6$ modes.
For $d=4$, with a maximum number of $n=7$ mode, we are able to match the fidelity up to $F(\rho_4,\tau_7) \approx 0.983$. 
After optimizing the measurement directions and binning choices, the state $\rho_7$ achieves a high CHSH score of $S \approx 2.34$.

\begin{figure}
    \centering
    \includegraphics[width=1\linewidth]{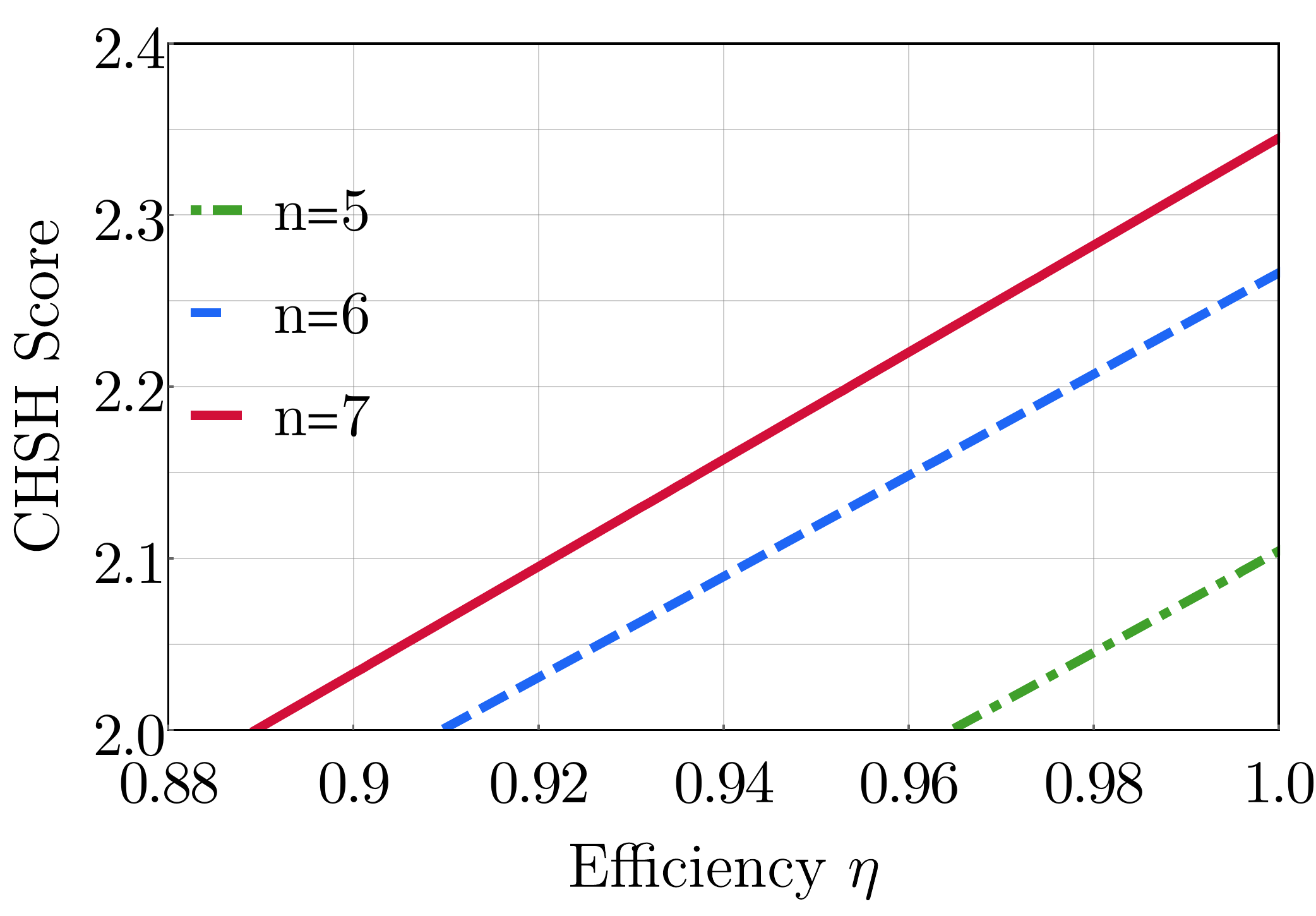}
    \caption{CHSH Score optimized over measurement and binning choices for state $\tau_n$ obtained by maximizing $F(\rho_{d=4},\tau_n)$. Results are given for circuits with 5,6 and 7 bosonic modes.} 
    \label{fig:robustnessd4}
\end{figure}

Finally, we study the robustness to losses for some of these state preparations.
In the $d=3$ case, considering the $n\geq 6$-modes circuits, we recovered the results obtained using the optimization of Eq~\eqref{eq:GenOptimNoise}. 
In Fig.~\ref{fig:robustnessd4}, we show the evolution of the CHSH score optimized over binning and measurement choices for the state prepared with circuit targeting the $\rho_4$ state.

%


\section*{Conclusion}
\label{sec:conclusion}

In this work we consider the task of violating Bell inequalities using homodyne measurements.  In local qubit Fock spaces, we develop  a method that suggests the CHSH inequality is violated only in the space spanned by 0 and 2 photons. Then we consider local qudit Fock spaces of increasing dimensions and optimize the CHSH score in each case. We found states that yield the largest violations observed in the literature for a fixed dimension. We focus on the experimental feasibility, first by deriving thresholds of efficiency required to enable Bell inequalities violations, then by proposing realistic set-ups which can produce such violations. We believe that this represents a significant step towards a Bell experiment with homodyne measurements. Future works include considering POVMs of more than $2$ outcomes, extending our results to the cases of heterodyne measurements and exploring the feasibility of device-independent protocol, including DIQKD, with homodyne measurements. 
\medbreak

\begin{acknowledgments}
This work was supported by the FastQI grant funded by the Institut Quantique Occitan.
E.O acknowledges support from the Government of Spain (FIS2020-TRANQI and Severo Ochoa CEX2019-000910-S), Fundació Cellex, Fundació Mir-Puig, Generalitat de Catalunya (CERCA, AGAUR SGR 1381) and from the ERC AdGCERQUT.
X.V. acknowledges funding by the European Union’s Horizon Europe research and innovation program under the project “Quantum Security Networks Partnership” (QSNP, Grant Agreement No. 101114043) and by a French national quantum initiative managed by Agence Nationale de la Recherche in the framework of France 2030 with the reference ANR-22-PETQ-0009.
\end{acknowledgments}


\bigskip
\setlength{\bibsep}{1pt plus 1pt minus 1pt}
\bibliography{biblio}


\clearpage

\appendix

\section{2-outcomes POVM for qubits implemented with homodyne measurement}
\label{app:sec2outcome}

\paragraph{Qubit POVM}
A Positive operator-valued measured (POVM) with k outcomes is a set of Hermitian operator $\{E_i\}_{i=1}^k$ which are positive $E_i\geq 0~~\forall~~i$ and satisfy the normalisation condition $\sum_i E_i=I.$ In the case of qubits one can write an arbitrary Hermitian operator in the Bloch form:
\begin{equation}
    E_i=\lambda_i(\mathds{1}+\Vec{n}_i.\Vec{\sigma})
\end{equation}
where $|\Vec{n_i}|\leq1$
In the case of 2-outcomes the normalisation condition impose that $\sum_i \lambda_i\Vec{n_i}=0$ thus $ \Vec{n_1}=-\frac{|\Vec{n_2}|}{|\Vec{n_1}|}\Vec{n_2}$ and $\lambda_1+\lambda_2=1.$ One can rewrite such POVM as a convex mixture of extremal POVM:
\begin{equation}
\label{povm}
\begin{split}
    E_1 &= \mu \ket{\vec{n}}\bra{\Vec{n}}+(1-\mu) r_1 \mathds{1}, \\
    E_2 &= \mu \ket{\vec{-n}}\bra{\Vec{-n}}+(1-\mu) r_2 \mathds{1}   
\end{split}
\end{equation}

where $\ket{\vec{n}}\bra{\Vec{n}} = \vec{n}.\vec{\sigma}$,   $\mu=2|n_i|\lambda_i$ and $r_i=\frac{(\lambda_i-\mu)}{1-\mu}.$ $\mu$ represent the projective part of the POVM, $(1-\mu)$ the random part and $\ket{\vec{\pm n}}$ is the eigenvector associated to the $\pm 1$ eigenvalue of $\vec{n_i}.\vec{\sigma}$. \\
\paragraph{Optimizing the trace distance}
The trace distance $\mu_h$ defined in section~\ref{sec:witness} reads:

\begin{equation}
\mu_h= \frac{1}{2}\int ||p(\theta,\ket{+\vec{n}}-p(\theta,\ket{-\vec{n}})||.
\end{equation}
The maximum of $\mu_{H}$ is achieved when $D(x,m,l,\phi,a,\theta)=||p(\theta,\ket{+\vec{n}}-p(\theta,\ket{-\vec{n}})||$ is maximum, the explicit expression of $D(x,m,l,\phi,a,\theta)$ being
\begin{equation}
\label{traceexpres}
\begin{split}
     D(&x,m,l,\phi,a,\theta) = \\
   & \bigg| \bigg| e^{-x^2} \biggl(\frac{2 \sin (2 a) H_l(x) H_m(x) \cos (\theta  m-\theta  l+\phi )}{\sqrt{\pi } \sqrt{2^m
   m!} \sqrt{2^l l!}} \\
    &+\cos (2 a) \biggl(\frac{2^{-n} H_l(x){}^2}{\sqrt{\pi } l!}-\frac{2^{-m}
   H_m(x){}^2}{\sqrt{\pi } m!}\biggr)\biggr) \bigg| \bigg|
\end{split}
\end{equation}
Where the states  $\ket{\pm n}$  are parametrized as
\begin{eqnarray}
     \ket{+\vec{n}}=\cos(a)\ket{l}+e^{i \phi}\sin(a)\ket{m}\\
     \ket{-\vec{n}}=\cos(a)\ket{l}-\sin(a)e^{i \phi}\ket{m}.
\end{eqnarray}
The maximum value of  $D(x,m,l,\phi,a,\theta)$ over $\phi,~a$ and $\theta$ is necessarily achieved when  $|\cos (\theta  m-\theta  l+\phi )|=1$, so $\theta  m-\theta  l+\phi=0$ or $\pi,$. The first case corresponds to $\theta=0$ and $\phi=0$ and the second case $\pi$ can be brought back to the first one by changing $a$ into $-a$ in \eqref{traceexpres}. Therefore, we have that:
\begin{equation}
\label{optitrace}
  \max_{\phi,a,\theta}  \mu_{H}= \max_{\phi=0,a,\theta=0}\mu_{H}.
\end{equation}
The worst case scenario, in order to violate the CHSH inequality with measurement of the form~\eqref{povm} is to have $r_1+r_2=\frac{1}{2}$ in this case we use the correlator 
\begin{equation}
    \sigma_{\ket{+\vec{n}}}=\mu(\ket{\vec{n}}\bra{\Vec{n}}-\ket{-\vec{n}}\bra{-\Vec{n}}).
\end{equation}
We consider the case where Alice and Bob are using the same settings, the corresponding CHSH operator 
\begin{equation}
\label{bellopapen}
\begin{split}
    \mathds{B}= &\sigma_{\ket{+\vec{n}_1}}\sigma_{\ket{+\vec{n}_1}} \\
    +&\sigma_{\ket{+\vec{n}_1}}\sigma_{\ket{+\vec{n}_2}} \\
    +&\sigma_{\ket{+\vec{n}_2}}\sigma_{\ket{+\vec{n}_1}} \\
    -&\sigma_{\ket{+\vec{n}_2}}\sigma_{\ket{+\vec{n}_2}}
\end{split}
\end{equation}
The maximum value of the CHSH inequality is achieved when measure with the operator \eqref{bellopapen} is obtain where the Bloch vector of $\ket{+\vec{n}_1}$ and $\ket{+\vec{n}_2}$ are orthogonal. We showed that the maximun of $\mu_{h}$ is achieved for a real state $\ket{+\vec{n}}.$ Let $a_{max}$ be the optimal value of $a$ for the optimization \eqref{optitrace}. The corresponding state is 
\begin{equation}
    \ket{+\vec{n}_1}=\text{cos}(\frac{a_{max}}{2})\ket{l}+\text{sin}(\frac{a_{max}}{2})\ket{m}.
\end{equation}
We write $\ket{+\vec{n}_2}$ as
\begin{equation}
\ket{+\vec{n}_2}=\text{cos}(\frac{a_{2}}{2})\ket{l}+e^{i\phi_2}\text{sin}(\frac{a_{2}}{2})\ket{m}.
\end{equation}
 One can see from \eqref{traceexpres} that one can always chose  $\theta$ such that $D(x,m,l,\phi,a,\theta)=D(x,m,l,0,a,0)$ we then choose $a_2=a_max$ and get $\mu_{\vec{n}_2}=\mu_{\vec{n}_1}$ for all $\phi_2.$ Finally the condition on $a_{max}$ for which it exist a $\phi_2$ such that the Bloch vector of $\ket{+\vec{n}_2}$ and $\ket{+\vec{n}_1}$ are orthogonal is $|\text{cos}(a_max)^2\leq\frac{1}{2}.$Therefore for all $\mu_{\ket{\vec{n}}}$ such that $|\text{cos}(a)^2\leq\frac{1}{2}$ One can achieve the CHSH violation of 
\begin{equation}
S=\mu^2 2\sqrt{2}.
\end{equation}

\paragraph{Certification of best possible quantum bound}
In order to obtain the best possible score to a Bell inequality, the states retained should be as distinguishable as possible, which means that their overlap should be as small as possible. With previous notations, considering two orthogonal states, $\mu$ can be optimized so that it exactly quantifies the overlap between them. Besides, there is a threshold value $\mu_t$ that certifies that the local bound can be exceeded. But for a fixed value of $\mu$, the values $r_i$ can be optimized also. Loosely speaking, they represent the best way to construct a POVM, taking into account the difference between the values of the two probability densities within the area where they overlap. In order to formalize these ideas, let us consider the two following probability densities:
\begin{equation}
    f(x) = |\bra{x}\ket{\vec{n}}|^2, \quad g(x) = |\bra{x}\ket{\vec{-n}}|^2
\end{equation}
where $\ket{\vec{n}}$ and $\ket{\vec{-n}}$ have been defined just above. Let us consider the POVM associated to an homodyne detection. We can write once again in two different ways, in the $\{ \ket{\vec{n}},\ket{\vec{-n}} \}$ basis.
\begin{equation}
\begin{split}
     E_1 &= \begin{bmatrix}
\int_{A} f(x) & 0 \\
0 & \int_{A} g(x)
\end{bmatrix} \\
    &= \begin{bmatrix}
\mu + (1 - \mu) r_1 & 0 \\
0 & (1 - \mu) r_1
\end{bmatrix}
\end{split}
\end{equation}
where $A$ is the interval chosen for the binning. It yields
\begin{equation}
    \mu = \int_A f-g, \quad r_1 = \frac{\int_A g}{1 - \int_A \left(  f-g \right)}
\end{equation}
Finally, in order to prove that the best numerical result is always obtained by an optimization on the projective part only, we have to show that 
the value of the random part $r_1$ is controlled, that is upper bounded, whenever the projective part $\mu$ is above a certain value. The final problem is written:
\begin{align*}
\optproblem{A} & \ \frac{\int_{A} g}{1 - \int_{A} (f-g)} \\
\text{s.t.} & \quad \int_{A} ( f-g ) > \mu_c \\
            & \quad f,g \in \mathcal{H}_2
\end{align*}
where $\mathcal{H}_2 = \{ |\lambda_{n_1} H_{n_1}(x) + \lambda_{n_2} H_{n_2}(x)|^2 | \lambda_{n_1} + \lambda_{n_2} = 1 \}$. 


\section{Noise Model for operators}
\label{App:NoiseModel}
In this section, we propose to test the robustness of the violations with regard to experimental conditions. The most common source of noise in these experiments are photon losses, which can be modelled by the action of a beam-splitter which entangles an ideal incoming state with an ancillary fluctuating quantum field, in that case the void state $\ket{0}$. After recombination on the beam-splitter, two outputs are produced corresponding to the transmitted part of the beam-splitter and to the reflected one. In order to obtain the noisy operator, we trace-out the part corresponding to the reflection. Hence, stemming from the ideal observable:
\begin{equation}
      \hat{\mathcal{O}}= +1\int_E \text{d}X\ket{X}\bra{X}-1\int_{\overline{E}} \text{d}X\ket{X}\bra{X}
\end{equation}
our final goal is to compute:
\begin{equation}
    \bra{0} \hat{U} \ \hat{\mathcal{O}} \ \hat{U}^{\dagger} \ket{0}
    \label{eq:NoisyOp}
\end{equation}
where $\hat{U}$ is the beam-splitter observable defined by $ \hat{U} = e^{i \theta (\hat{a}^\dagger \hat{b} - \hat{a} \hat{b}^\dagger)} $, $\hat{a}, \hat{b}$ are the annihilation operators respectively on the first and second mode, and the reflection is $\eta = \cos(\theta)^2$. 
We use
\begin{equation}
    \ket{X}\bra{X} = \frac{1}{2 \pi} \int_{\mathds{R}} e^{i \xi (\hat{x} - x)}
\end{equation}
and
\begin{equation}
    \hat{x} = \frac{\hat{a} + \hat{a}^\dagger}{\sqrt{2}} 
\end{equation}
and with $\hat{A} = i \theta (\hat{a}^\dagger \hat{b} - \hat{a} \hat{b}^\dagger)$ and $\hat{B} = \frac{i \xi (\hat{a} + \hat{a}^\dagger) }{\sqrt{2}}  $ the quantity to compute is now:
\begin{equation}
\int_I \text{d}X\ket{X}\bra{X} = \\
    \int_I \frac{\text{d}x}{2 \pi} \int e^{-i \xi x}\text{d}\xi e^{\hat{A}} e^{\hat{B}} e^{-\hat{A}}.
    \label{eq:BSopintegral}
\end{equation}
We are going to first develop the product of the exponential of operators under the integral. To begin with, let us note that, with $\text{K}_\xi = \frac{i \xi}{\sqrt{2}}$
\begin{equation}
    [\text{K}_\xi \hat{a}, \text{K}_\xi \hat{a}^\dagger] = \frac{\xi^2}{2} \mathds{1}
\end{equation}
and since this commutator commutes with $\text{K}_\xi \hat{a}$ and $\text{K}_\xi \hat{a}^\dagger$, the Baker-Campbell-Hausdorff formula yields : 
\begin{equation}
    \exp(\hat{B}) = \exp(\text{K}_\xi \hat{a}) \exp(\text{K}_\xi \hat{a}^\dagger) \exp(\frac{-\xi^2}{4}).
\end{equation}
We use 
\begin{equation}
    e^{\hat{A}} e^{-\hat{A}} = \mathds{1}
\end{equation}
to finally rewrite $e^{\hat{A}} e^{\hat{B}} e^{-\hat{A}}$ as:
\begin{equation}
\begin{split}
    \exp(\frac{-\xi^2}{4}) &\exp(\hat{A}) \exp(\text{K}_\xi \hat{a})\exp(-\hat{A})\\
    \times &\exp(\hat{A}) \exp(\text{K}_\xi \hat{a}^\dagger)   \exp(-\hat{A})
\end{split}
     \label{eq:expoperatordec}
\end{equation}
The Campbell identity yields: 
\begin{equation}
    \exp(\hat{A}) \exp(\text{K}_\xi \hat{a})   \exp(-\hat{A}) =  \sum_{n=0}^{\infty} \frac{[(\hat{A})^n,\text{K}_\xi \hat{a}]}{\text{n!}}
\end{equation}
where $[(\hat{A})^n,\text{K}_\xi \hat{a}] = [\hat{A},...,[\hat{A},[\hat{A},\text{K}_\xi \hat{a}]]]$.
We separate this sum into its even and its odd parts, and use the fact that:
\begin{equation}
    [(\hat{A})^{2p},\text{K}_\xi \hat{a}] = \text{K}_\xi (i \theta)^{2 p } \hat{a}
\end{equation}
and 
\begin{equation}
    [(\hat{A})^{2p+1},\text{K}_\xi \hat{a}] = \text{K}_\xi (i \theta)^{2 p +1} \hat{b}
\end{equation} to finally obtain:
\begin{equation}
    \begin{split}
     \exp(\hat{A}) & \exp(\text{K}_\xi \hat{a})\exp(-\hat{A}) \\
        &= \sum_{p \geq 0} \frac{(i \theta)^{2p}}{(2p)!} \hat{a} +  \sum_{p \geq 0} \frac{(i \theta)^{2p+1}}{(2p+1)!} \hat{b} \\
        &= \text{K}_\xi \left( \cosh{(i \theta)} \hat{a} + \sinh{(i \theta)} \hat{b} \right) \\
        &= \text{K}_\xi \left( \cos{(\theta)} \hat{a} + \sin{(\theta)} \hat{b} \right).       
    \end{split}
    \label{eq:CampbellBS}
\end{equation}
Now, using Eq.~\eqref{eq:CampbellBS} and Eq. ~\eqref{eq:expoperatordec}, we have that Eq.~\eqref{eq:BSopintegral} is equal to:
\begin{equation}
\begin{split}
        \int_I \frac{\text{d}x}{2 \pi} \int \text{d}\xi \Big( &e^{-i \xi x} \\
        \times &e^{\frac{-\xi^2}{4}} e^{\text{K}_\xi ( \cos{(\theta)} \hat{a}^\dagger + \sin{(\theta)} \hat{b}^\dagger )} \\ 
        \times &e^{ \text{K}_\xi ( \cos{(\theta)} \hat{a} + \sin{(\theta)} \hat{b} ) } \Big)
\end{split}
\end{equation}
Finally, we apply $\bra{0} \cdot \ket{0}$ and Eq.~\eqref{eq:NoisyOp2} becomes: 
\begin{equation}
        \int_E \frac{\text{d}x}{2 \pi} \int \text{d}\xi e^{-i \xi x} e^{\frac{-\xi^2}{4}} \exp^{\text{K}_\xi  \cos{(\theta)} \hat{a}^\dagger } \exp^{ \text{K}_\xi  \cos{(\theta)} \hat{a} }
\end{equation}
For any $(k,n) \in \mathds{N}^2, k \leq n$, 
\begin{equation}
    \hat{a}^k \ket{n} = \sqrt{(n)} \sqrt{(n-1)}...\sqrt{(n-k-1)} \ket{n-k}
\end{equation}
So
\begin{equation}
\begin{split}
    \exp &(\text{K}_{\xi}  \cos{(\theta)} \hat{a}) \ket{n} \\
    &\quad= \sum_{k=0}^\infty \frac{\left( \text{K}_\xi  \cos{(\theta) \hat{a}} \right)^k}{k!} \ket{n} \\
   &\quad= \sum_{k=0}^n \frac{\left( \text{K}_\xi  \cos{(\theta)} \right)^k \sqrt{\binom{n}{k}} \ket{n-k} }{\sqrt{k!}} 
\end{split}
\end{equation}


\section{Realistic implementation of Bell violation with homodyne measurements}
\label{app:Circuits}

\begin{figure}[t]
	\begin{center}
		\includegraphics[width=0.45\textwidth]{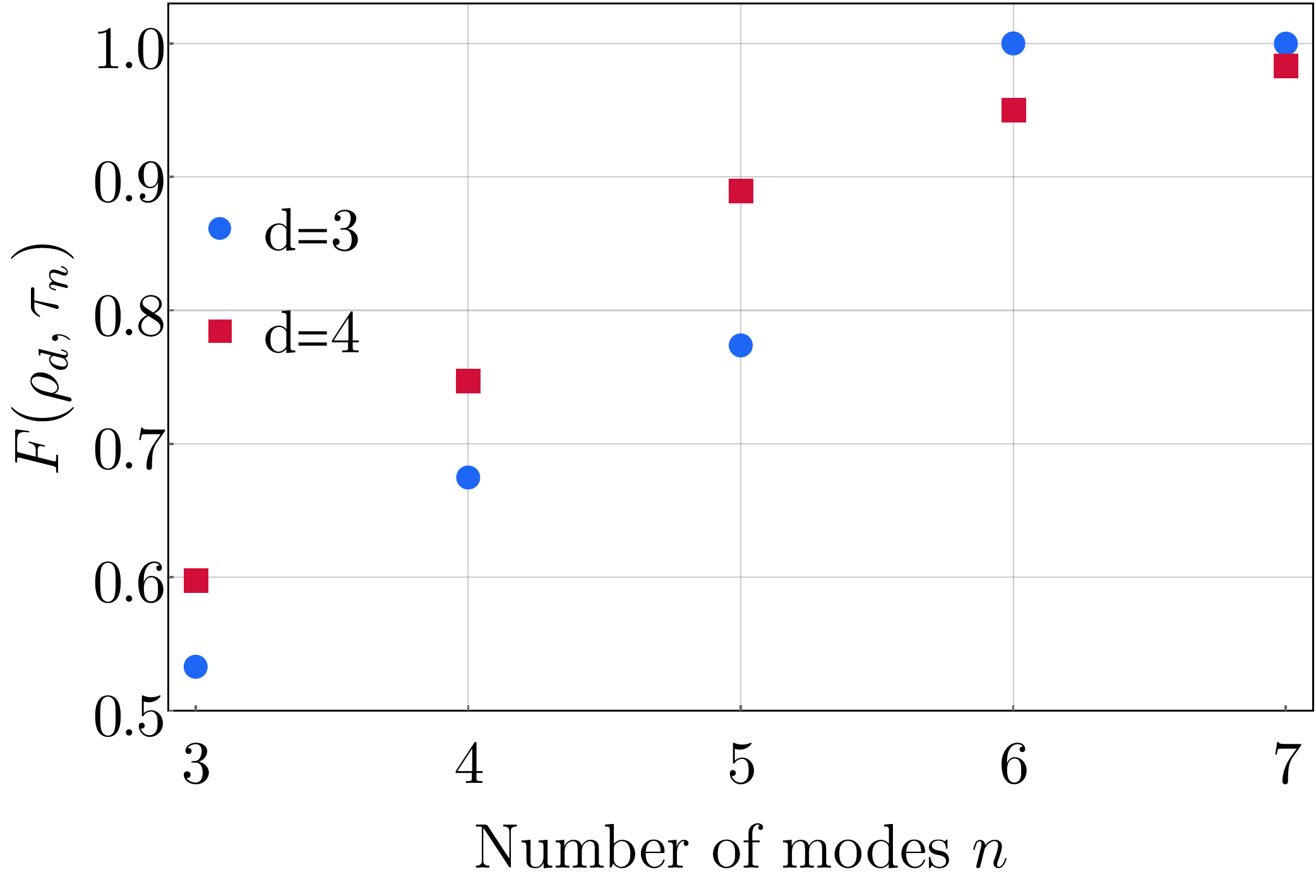}
	\end{center}
	\caption{Optimized fidelities $F(\rho_d,\tau_n)$ given for local dimensions $d=3$ and $d=4$ and for circuits of $n=\{3,4,5,6,7\}$ modes circuits.}\label{fig:fidelities}
\end{figure}

\begin{figure}[t]
	\begin{center}
		\includegraphics[width=0.45\textwidth]{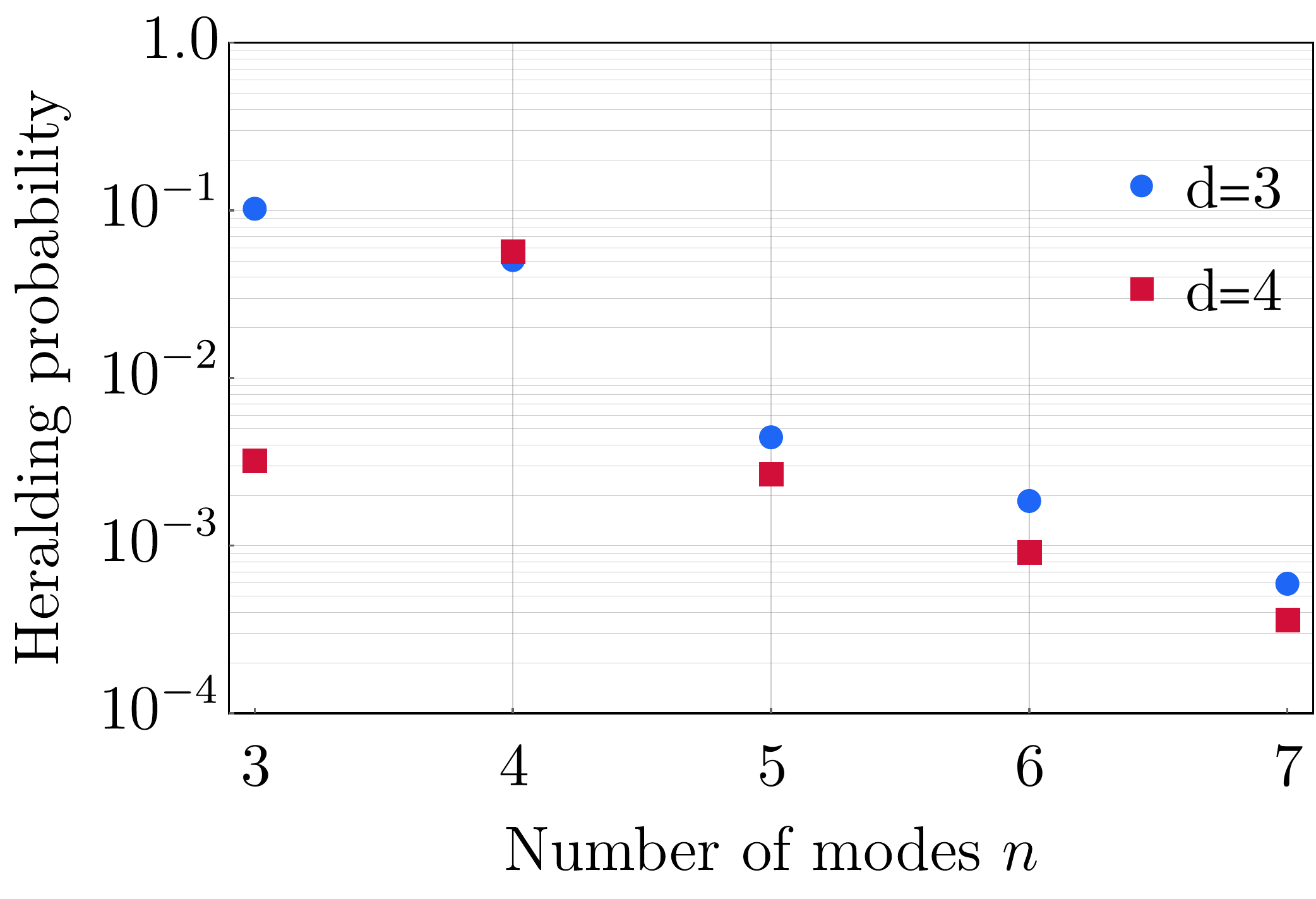}
	\end{center}
	\caption{Heralding probability for circuits achieving the best fidelities for local dimensions $d=3$ and $d=4$ and using $n=\{3,4,5,6,7\}$ modes.}\label{fig:heralding}
\end{figure}

To study the feasibility of implementing states yielding a high CHSH score, we consider circuits using one Gaussian operation $\mathcal{G}$ with zero-displacement followed by displacement operations, as shown in Fig.~\ref{fig:circuit}.
For local dimension $d=3,4$ and for ciruits up to $n=7$ modes, we give the optimized fidelity $F(\rho_d,\tau_n)$ over circuits parameters in Fig.~\ref{fig:fidelities}.
For these circuits, we also provide in Fig.~\ref{fig:heralding} the heralding probability, i.e. the probability to obtain a detection event on all $n-2$ last modes simultaneously.
In order to better understand the experimental resources needed for the proposed implementations, we here provide optimal circuit parameters in terms of squeezing and displacement amplitudes.

\smallbreak

For clarity, we first define squeezing and displacement operations.
Single-mode squeezing operations on mode $i$ act following
\begin{equation}
    S(z) = \exp\left(\frac{1}{2}(z^* \hat{a}_i^2 -z (\hat{a}_i^\dagger)^2)\right),
\end{equation}
where $\hat{a}_i$ and $\hat{a}_i^\dagger$ are the ladder operators of mode $i$, and $z=r\exp(i\phi)$ with $r\in\mathds{R}$ and $\phi\in[0,2\pi]$.
Displacement operations on mode $i$ are given by
\begin{equation}
    D(\alpha) = \exp\left(\alpha \hat{a}_i -\alpha^* \hat{a}_i^\dagger\right),
\end{equation}
with $\alpha\in\mathds{C}$.

\smallbreak

To obtain the squeezing parameter from the unitary $\mathcal{G}$ we use the Bloch-Messiah decomposition~\cite{Bloch1962,Braunstein2005}.
In phase-space, the transformation $\mathcal{G}$ is fully characterized by a symplectic matrix $\Omega \in \mathds{R}^{2n \times 2n}$.
The Bloch-Messiah decomposition implies that $\Omega$ can be written as $\Omega = O_1 Z O_2$, with $O_1,O_2$ two orthogonal symplectic matrices, and $Z$ a diagonal matrix.
Moreover, when acting on the vacuum, as it is the case in setups we consider, the orthogonal property of $O_1$ and $O_2$ allow us to simplify the expression to
\begin{equation}
    \mathcal{G} = O_1 Z.
\end{equation}
$O_1$ can be interpreted as passive Gaussian transformations, i.e. as a combination of phase-shifters and beam-splitters, while $Z=\text{diag}(\exp(-r_1),\dots,\exp(-r_n),\exp(r_1),\dots,\exp(r_n))$ represents an array of single-mode squeezer acting with parameter $z=r_i$ on mode $i$.

We use the Bloch-Messiah decomposition implemented in \textsc{StrawberryFields}~\cite{Killoran2019}.
In Table \ref{tab:l=3}, we give the optimal squeezing and displacement parameters when optimizing $\rho_3$ for circuits up to $n=7$-modes.
These parameters are given for circuits targeting $\rho_4$ in Table \ref{tab:l=4}.

\smallbreak

While no constraints are set on the squeezing parameters, we found a maximum squeezing of $r\approx 2.3$. Using
\begin{equation}
    V_{\text{dB}} = -10\times\log_{10}(\exp(-2r)),
\end{equation}
this corresponds to $20$dB squeezed vacuum states. 
Importantly, the $n=6$-modes circuit preparing exactly the state $\rho_{d=3}$ only requires $r\leq 1.485$ or $12.9$dB of squeezing.
In the $d=4$ case, the $7$-mode photonic circuits needs $r\leq 1.435$ or $12.5$dB squeezed vacuum states. 
This is well below experimental limits, as direct observations of up $15$dB squeezing have been reported \cite{Vahlbruch2016}. 
Note that more refined approaches to design photonic circuits, e.g. see \cite{Valcarce2023,Krenn2020,RuizGonzalez2023,Melnikov2020}, might be necessary to better match or ease other experimental constraints for the preparation of the proposed quantum states. More specifically, such method could be used in combination with the optimisation ~\eqref{eq:GenOptimbinning} to directly target a high CHSH score.

\begin{table}[t]
\begin{tabular}{l || c | c | c| c| c | c | c||}
	mode & 1 & 2 & 3 & 4 & 5 & 6 & 7\\
	\hline
	\hline
 & \multicolumn{7}{c||}{Squeezing parameter $r$} \\
 \hline
	n=3 & 0.6872 & 0.6631 & 0.3934 & & & & \\
	n=4 & 1.2387 & 1.1164 & 0.6693 & 0.5484 & & & \\
	n=5 & 2.2971 & 1.3560 & 1.1815 & 0.4920 & 0.2930 & & \\
	n=6 & 1.4834 & 1.2309 & 0.7445 & 0.6903 & 0.5410 & 0.2405 & \\
	n=7 & 1.1953 & 1.1670 & 0.7855 & 0.6586 & 0.6285 & 0.3194 & 0.2701 \\
\hline
\hline
 & \multicolumn{7}{c||}{Displacement $d$} \\
 \hline
	n=3 & & & 0.5089 & & & & \\ 
	n=4 & & & 0.3527 & -0.3077 & & & \\ 
	n=5 & & & 0.7174 & -0.4598 & 0.4729 & & \\ 
	n=6 & & & 0.4847 & -0.4373 & 0.4958 & -0.4986 & \\ 
    n=7 & & & 0.4781 & -0.5003 & 0.4906 & -0.4955 & 0.4785 \\ 
\end{tabular}
\caption{Circuits parameters maximizing the fidelity $F(\rho_3, \tau_n)$ when considering $n={3,4,5,6,7}$ bosonic modes. }
\label{tab:l=3}
\end{table}

\begin{table}
\begin{tabular}{l || c | c | c| c| c | c | c||}
	mode & 1 & 2 & 3 & 4 & 5 & 6 & 7\\
	\hline
	\hline
 & \multicolumn{7}{c||}{Squeezing parameter $r$} \\
 \hline
	n=3 & 2.0618 & 0.6903 & 0.1571 & & & & \\
	n=4 & 1.1707 & 1.1027 & 0.8939 & 0.3295 & & & \\
	n=5 & 1.6180 & 1.3506 & 0.9420 & 0.6105 & 0.3116 & & \\
	n=6 & 1.7885 & 1.0188 & 0.9710 & 0.8686 & 0.5690 & \\
	n=7 & 1.4353 & 1.0714 & 0.8333 & 0.6945 & 0.6392 & 0.2438 \\
\hline
\hline
 & \multicolumn{7}{c||}{Displacement $d$} \\
 \hline
	n=3 & & & 0.3862 & & & & \\ 
	n=4 & & & 0.4927 & -0.5405 & & & \\ 
	n=5 & & & 0.4326 & -0.4696 & 0.4467 & & \\ 
	n=6 & & & 0.4959 & -0.5035 & 0.5029 & -0.5790 & \\ 
    n=7 & & & 0.4137 & -0.4120 & 0.4019 & -0.5450 & 0.4990 \\ 
\end{tabular}
\caption{Circuits parameters maximizing the fidelity $F(\rho_4, \tau_n)$ when considering $n={3,4,5,6,7}$ bosonic modes. }
\label{tab:l=4}
\end{table}

\vfil


\end{document}